\documentstyle[11pt]{article}
\def \qed{\hfill
    $\vrule height 2.5mm
width 2.5mm depth 0mm $}
\newtheorem{lem}{Lemma}
\newtheorem{th}{Theorem}
\newtheorem{pr}{Proposition}
\newtheorem{col}{Corollary}
\begin{document}
\title{Unimodality of generalized Gaussian coefficients.}
\vskip 0.4cm
\author{\Large {Anatol N. Kirillov} \\
{\small {\it Steklov Mathematical Institute,}} \\
{\small {\it Fontanka 27, St.Petersburg, 191011,
Russia}}}
\date{January 1991}
\maketitle
\begin{abstract}
A combinatorial proof of the unimodality of the generalized  \linebreak
$q$-Gaussian
coefficients $\left[\begin{array}{c}N\\ \lambda \end{array}\right]_q$
based on the explicit formula for \linebreak Kostka-Foulkes polynomials
is given.
\end{abstract}
\vskip 0.5cm

 ${\bf 1^0.}$ \ Let us mention that the proof of the unimodality of the
generalized
Gaussian coefficients based on theoretic-representation considerations
was given by E.B. Dynkin \cite{d} (see also \cite{h}, \cite{s2}, \cite{s3}).
Recently
K.O'Hara \cite{oh} gave a constructive proof of the unimodality of the
Gaussian coefficient
$ \left[\begin{array}{c}
n+k\\n
\end{array} \right]_q=s_{(k)}(1,\cdots ,q^k)$,
and D. Zeilberger  \cite{z} derived some identity which may be consider as
an ``algebraization'' of O'Hara's construction. By induction this identity
immediately implies the unimodality of $\left[\begin{array}{c}
n+k\\n
\end{array} \right]_q $.
Using the observation (see Lemma~ 1) that the generalized Gaussian coefficient
$\left[\begin{array}{c}
n\\ {\lambda}'
\end{array} \right]_q $
may be identified (up to degree $q$ ) with the Kostka-Foulkes polynomial
$ K_{\widetilde{\lambda},\mu}(q)$ (see Lemma~ 1), the proof of the unimodality
of $\left[\begin{array}{c}
n \\ {\lambda}'
\end{array} \right]_q$
is a simple consequence of the exact formula for Kostka-Foulkes polynomials
contained in \cite{k2}. Furthermore the expression for
$K_{\widetilde{\lambda},\mu}(q)$ in the case $\lambda =(k)$  coincides  with
identity
(KOH) from \cite{8}. So we obtain a generalization and a combinatorial proof
of (KOH) for arbitrary partition $\lambda$.
\vskip 0.5 cm

 ${\bf 2^0.}$ \ Let us recall some well known facts which will be used later.
We base ourselves
\cite{9} and \cite{m}. Let $\lambda =({\lambda}_1\ge {\lambda}_2\ge \cdots)$
be a partition, $|\lambda |$ be the sum of its parts $\lambda _i$,
$n(\lambda)=\sum _i(i-1)\lambda _i$ and $\left[\begin{array}{c}
n \\ \lambda
\end{array} \right]_q $
be the generalized Gaussian coefficient.

Recall that
$$ s_{\lambda}(1,\cdots ,q^n)=q^{n(\lambda)}\left[\begin{array}{c}
n \\ \lambda '
\end{array} \right]_q=\prod _{x\in \lambda}\frac{1-q^{n+c(x)}}{1-q^{h(x)}}.
\eqno (1)$$
Here $c(x)$ is the content and $h(x)$ is the hook length corresponding
to the box
$x\in \lambda $, \cite{m}.

\begin{lem}
\ Let $\lambda$ be a partition and fix a positive integer $n$.
Consider new
partitions $ \widetilde{\lambda}=(n\cdot |\lambda |,\lambda )$ and
$\mu =(|\lambda |^{n+1})$. Then
$$ q^{\frac{n(n-1)}{2}\cdot |\lambda|+n(\lambda)}\left[\begin{array}{c}
n \\ \lambda '
\end{array} \right]_q=K_{\widetilde{\lambda},\mu}(q). \eqno (2)
$$
\end{lem}
Proof. We use the description of the polynomial $ q^{|\lambda|+n(\lambda)}
\cdot\left[\begin{array}{c}
n \\ \lambda '
\end{array} \right]_q$ as a generating function
for the standard Young tableaux of the shape $\lambda$ filled with
numbers from the interval $[1,\cdots ,n]$. Let us denote this set of Young
tableaux by $STY(\lambda ,\le n)$. Then
$$ q^{|\lambda|+n(\lambda)}\left[\begin{array}{c}
n \\ \lambda '
\end{array} \right]_q=\sum _{T\in STY(\lambda ,\le n)}q^{|T|}\ . \eqno (3)
$$
Here $|T|$ is the sum of all numbers filling the boxes of $T$ .
For any tableau $T$ (or diagram $\lambda$) let us denote by $T[k]$
(or $\lambda[k])$ the part of $T$ (or $\lambda$) consisting of rows
starting from the $(k+1)$-st \ one. Given tableau \
$T\in STY(\lambda ,\le n)$,
then consider tableau $\widetilde T\in STY(\widetilde{\lambda},\mu)$ such that
$\widetilde T[1]=T+supp \ \lambda [1]$, and we fill the first row of
$\tilde T$ with all remaining numbers in increasing order from left
to right. Here for any diagram $\lambda$
we denote by $ supp \lambda$ the plane partition of the shape $\lambda$
and content $(1^{|\lambda |})$.
It is easy to see that
$$
c(\widetilde T)=|T|+\frac{(n+1)(n-2)}{2}\cdot|\lambda|\ ,
$$
so we obtain the identity (2).
\qed

Let us consider an explanatory example. Assume $\lambda =(2,1) ,\ n=3$.
Then $\widetilde \lambda =(9,2,1), \ \mu=(3^4)$.
It is easy to see that $|STY(\lambda ,\le 3)|=8$.
\vskip 0.5cm
\begin{tabular}{ccccccccl}
 $ T$ &&& $ |T|$ && ${\widetilde T}$ &&& $ c(\widetilde T)$ \\
\hline
&&&&&&&&\\
$
\begin{array}{ll}
1 & 1 \\ 2
\end{array} $
&&& 4 && $
\begin{array}{lllllllll}
1&1&1&2&3&3&4&4&4\\2&2\\3
\end{array} $ &&&10\\
&&&&&&&&\\
$ \begin{array}{ll}
1&1\\3
\end{array}$&&&5&&
$\begin{array}{lllllllll}
1&1&1&2&3&3&3&4&4\\2&2\\4
\end{array}$ &&&11\\
&&&&&&&&\\
$\begin{array}{ll}
1&2\\2
\end{array}$&&&5&&
$\begin{array}{lllllllll}
1&1&1&2&2&3&4&4&4\\2&3\\3
\end{array}$ &&&11\\
&&&&&&&&\\
$\begin{array}{ll}
1&2\\3
\end{array}$ &&&6&&
$\begin{array}{lllllllll}
1&1&1&2&2&3&3&4&4\\2&3\\4
\end{array}$ &&&12\\
&&&&&&&&\\
$\begin{array}{ll}
1&3\\2
\end{array}$ &&&6&&
$\begin{array}{lllllllll}
1&1&1&2&2&3&3&4&4\\2&4\\3
\end{array}$ &&&12\\
&&&&&&&&\\
$\begin{array}{ll}
1&3\\3
\end{array}$ &&&7&&
$\begin{array}{lllllllll}
1&1&1&2&2&3&3&3&4\\2&4\\4
\end{array}$ &&&13\\
&&&&&&&&\\
$\begin{array}{ll}
2&2\\3
\end{array}$ &&&7&&
$\begin{array}{lllllllll}
1&1&1&2&2&2&3&4&4\\3&3\\4
\end{array}$ &&&13\\
&&&&&&&&\\
$\begin{array}{ll}
2&3\\3
\end{array}$ &&&8&&
$\begin{array}{lllllllll}
1&1&1&2&2&2&3&3&4\\3&4\\4
\end{array}$ &&&14
\end{tabular}
\vskip 0.2cm
Now we would like to use the formula for Kostka-Foulkes polynomials, obtained
in
\cite{k2}.
\vskip 0.5cm

 ${\bf 3^0.}$ \ First let us recall some definitions from \cite{k2}. Given
a partition $\lambda$ and composition $\mu$,
a configuration $\{\nu\}$  of the type $(\lambda,\mu)$
is, by definition, a collection of partitions
\ $\nu ^{(1)},\nu ^{(2)},\cdots $ such that

1)\enspace $ |\nu ^{(k)}|=\sum _{j\ge k+1}\lambda_j $;

2)\enspace $P_n^{(k)}(\lambda ,\mu):=Q_n(\nu ^{(k-1)})-2Q_n(\nu^{(k)})+
Q_n(\nu^{(k+1)})\ge 0$ for all $k,n\ge 1$,
\vskip 0.1cm

\hskip 0.6cm   where
\ $Q_n(\lambda ):=\sum _{j\le n}{\lambda}'_j$ ,and $ \ \nu^{(0)}=\mu .$
\begin{pr}
\cite{k2} \ Let $\lambda$ be a partition and $\mu$ be a composition, then
$$
K_{\lambda ,\mu}(q)=\sum_{\{\nu \}}q^{c(\nu )}\prod _{k,n}
\left[\begin{array}{c}
P_n^{(k)}(\lambda ,\mu)+m_n(\nu ^{(k)}) \\ m_n(\nu^{(k)})
\end{array} \right]_q, \eqno (4)
$$
where the summation in (4) is taken over all configurations of
$\{\nu\}$ \ of the type
$(\lambda,\mu), \ m_n(\nu^{(k)})=(\nu^{(k)})'_n-(\nu^{(k)})'_{n+1}$.
\end{pr}

{}From Lemma 1 and Proposition 1 we deduce
\begin{th} Let $\lambda$ be a partition. Then
$$
\left[\begin{array}{c}
N \\ \lambda '
\end{array} \right]_q=\sum_{\{\nu\}}q^{c_0(\nu)}\prod_{k,n}
\left[\begin{array}{c}
P_n^{(k)}(\lambda ,N)+m_n(\nu ^{(k)}) \\ m_n(\nu^{(k)})
\end{array} \right]_q, \eqno (5)
$$
where the summation in (5) is taken over all collections $\{\nu\}$ of
partitions \  $\{\nu\}=\{\nu^{(1)},\nu^{(2)},\cdots \}$ such that

1)\ $|\nu^{(k)}|=\sum_{j\ge k}\lambda _j,\ k\ge 1, \ |\nu^{(0)}|=0,$
\vskip 0.2cm
2)\ $P_n^{(k)}(\nu,N):=n(N+1)\cdot\delta _{k,1}+Q_n(\nu^{(k-1)})-
2Q_n(\nu^{(k)})+Q_n(\nu^{(k+1)})\ge 0$, for all $k,n\ge 1$.
Here
$$
c_0(\nu)=n(\nu^{(1)})-n(\lambda)+\sum_{
k,n \geq 1 }
\left(\begin{array}{c}
\alpha_n^{(k)}-\alpha_n^{(k+1)}\\2 \end{array}\right),\enspace
\alpha_n^{(k)}:=(\nu^{(k)})'_n \eqno (6)
$$
and by definition \ $\left( \begin{array}{l}
\alpha \\ 2 \end{array}\right):=\frac{\alpha(\alpha -1)}{2}$ \
for any  \ $\alpha \in{\bf R}$.
\end{th}

The identity (5) may be consided as a generalization of the (KOH) - identity
(see \cite{8}) for arbitrary partition $\lambda$.
\begin{col}
The generalized $q$-Gaussian coefficient $\left[\begin{array}{c}
n \\ \lambda '
\end{array} \right]_q$ \ is a symmetric and unimodal polynomial of
degree \ $(N-1)|\lambda|-n(\lambda)$.
\end{col}

Proof.  \ First, it is well known (e.g. [10],[11]) that the product of
symmetric and unimodal polynomials is again symmetric and unimodal. Secondly,
we use a well known fact (e.g.[10]), that the ordinary $q$-Gaussian
coefficient \ $\left[\begin{array}{c}m+n\\n\end{array}\right]_q$ \ is a
symmetric and
unimodal polynomial of degree $mn$. So in order to prove Corollary 1, it is
sufficient to show that the sum
$$
2c_0(\nu)+\sum_{k,n}m_n(\nu^{(k)})P_n^{(k)}(\nu,N)  \eqno (7)
$$
is the same for all collections of partitions \ $\{\nu\}$ \ which satisfy
the conditions 1) and 2) of the Theorem 1.
In oder to compute the sum (7), we use the following result (see [4]):
\begin{lem}
Assume \ $\{\nu\}$ \ to be a configuration of the type \ $(\lambda,\nu)$.
Then
$$
\sum_{k,n}m_n(\nu^{(k)})P_n^{(k)}(\nu,\mu)=
2n(\mu)-2c(\nu)-\sum_{n\geq 1}\mu '_n
\cdot \alpha_n^{(1)} . \eqno (8)
$$
\end{lem}

Using Lemma 2, it is easy to see that the sum (7) is equal to \
${(N-1)}|\lambda|-n(\lambda)$. This concludes the proof.
\qed

Note that in the proof of  Corollary 1 we use symmetry and unimodality of
the ordinary $q$-Gaussian coefficient
$\left[\begin{array}{c}m+n\\n\end{array}\right]_q$.
However, we may prove the unimodality of
$\left[\begin{array}{c}m+n\\n\end{array}\right]_q$ \
by induction using the identity (5) for the case \ $\lambda =(1^n),\ N=m$.

{\bf Remark 1.}\enspace The unimodality of generalized $q$-Gaussian
coefficients was also proved in the recent preprint [7]. The proof in [7]
uses the result from [4]. However [7] does not contain the identity (5).

{\bf Remark 2.}\enspace The proof of the identity (4) given in [4] is based on
the construction and properties of the bijection (see [4])
$$
STY(\lambda,\mu)\rightleftharpoons QM(\lambda,\mu).
$$
It is an interesting task to obtain an analytical proof of (5).
In the case $q=1$
such a proof was obtained in [3].

{\bf Acknowledgements.}\ The final version of this paper was written during
the author's stay at RIMS. The author expresses his deep gratitude to RIMS
for its hospitality.

\end{document}